# Search For Unresolved Sources
# In The *COBE* [1]-DMR Two-Year Sky Maps

A. Kogut[2], A.J. Banday[3], C.L. Bennett[4], G. Hinshaw[2], K. Loewenstein[2],
P. Lubin[5], G.F. Smoot[6], and E.L. Wright[7]

astro-ph/9402007  2 Feb 1994

---

[1] The National Aeronautics and Space Administration/Goddard Space Flight Center (NASA/GSFC) is responsible for the design, development, and operation of the Cosmic Background Explorer (*COBE*). Scientific guidance is provided by the *COBE* Science Working Group. GSFC is also responsible for the development of analysis software and for the production of the mission data sets.

[2] Hughes STX Corporation, Code 685.3, NASA/GSFC, Greenbelt MD 20771

[3] Universities Space Research Association, Code 685.9, NASA/GSFC, Greenbelt MD 20771

[4] NASA Goddard Space Flight Center, Code 685, Greenbelt MD 20771

[5] UCSB Physics Department, Santa Barbara CA 93106

[6] LBL, SSL, & CfPA, Bldg 50-351, University of California, Berkeley CA 94720

[7] UCLA Astronomy Department, Los Angeles CA 90024-1562

# ABSTRACT

We have searched the temperature maps from the *COBE* Differential Microwave Radiometers (DMR) first two years of data for evidence of unresolved sources. The high-latitude sky ($|b| > 30°$) contains no sources brighter than 192 $\mu$K thermodynamic temperature (322 Jy at 53 GHz). The cumulative count of sources brighter than threshold $T$, $N(>T)$, is consistent with a superposition of instrument noise plus a scale-invariant spectrum of cosmic temperature fluctuations normalized to $Q_{rms-PS} = 17$ $\mu$K. We examine the temperature maps toward nearby clusters and find no evidence for any Sunyaev-Zel'dovich effect, $\Delta y < 7.3$ $\times 10^{-6}$ (95% CL) averaged over the DMR beam. We examine the temperature maps near the brightest expected radio sources and detect no evidence of significant emission. The lack of bright unresolved sources in the DMR maps, taken with anisotropy measurements on smaller angular scales, places a weak constraint on the integral number density of any unresolved Planck-spectrum sources brighter than flux density $S$, $n(>S) < 2 \times 10^4$ $(S/1 \text{ Jy})^{-2}$ sr$^{-1}$.

*Subject headings:* cosmic microwave background — radio continuum:general

## 1. Introduction

The *COBE*-DMR sky maps represent an unbiased survey of the full sky at frequencies 31.5, 53, and 90 GHz (wavelengths 9.5, 5.7, and 3.3 mm) and 7° angular resolution. Expected unresolved astrophysical sources at these wavelengths include Galactic HII regions, extragalactic radio and infrared sources, inverse Compton scattering of cosmic microwave background (CMB) photons by hot gas in clusters of galaxies (Sunyaev & Zel'dovich 1980), as well as CMB anisotropy from varying gravitational potentials on the surface of last scattering (Sachs & Wolfe 1967).

Bennett et al. (1992b, 1993) limit the total contribution from unresolved non-cosmic sources averaged over the high-latitude ($|b| > 30°$) sky by comparing the first-year DMR maps to maps of known source distributions. With the exception of the Galactic quadrupole, there is no evidence for significant non-cosmological contribution to the anisotropy in the DMR sky maps. Since these results depend on an average over a large area of the sky, they do not preclude detection of weak non-cosmological emission restricted to smaller regions of the sky. A pixel-by-pixel search for sources in the first year DMR data shows no evidence for unresolved sources at the 200 $\mu$K level (Bennett et al. 1993). We report here further results of a search for a population of weak unresolved sources in the two-year DMR maps, using both pixel-by-pixel techniques and a statistical comparison of the maps at positions where non-cosmological emission is expected to be strongest.

## 2. Full-Sky Search

The DMR by design is insensitive to unresolved sources. The instrument response $T_A$ to a source temperature distribution $T(\theta, \phi)$ is

$$T_A = \frac{\int d\Omega \ T(\theta, \phi) \ P(\theta, \phi)}{\int d\Omega \ P(\theta, \phi)}$$

where we approximate the inner portion of the antenna pattern $P(\theta, \phi)$ as a Gaussian with 7° full width at half-maximum. Sources with angular size smaller than the beam area (solid angle $\sim$ 69 square degrees, Bennett et al. 1992a) are diluted by the ratio of the source area to beam area. Flux densities 65, 180, and 536 Jy are required to produce a 100 $\mu$K signal in the 31, 53, and 90 GHz channels, respectively.

The DMR maps are dominated by instrument noise and nearly scale-invariant CMB anisotropy (Smoot et al. 1992, Bennett et al. 1994). The DMR beam is wider than the 2.6° diameter pixels used to map the temperature anisotropy. Unresolved sources in the DMR maps can be distinguished from noise peaks by their correlated signal in neighboring pixels. We combine the two channels, A and B, at each DMR frequency to form "sum" (A+B)/2 and "difference" (A-B)/2 maps. For each pixel of these maps, we subtract the mean of all other pixels within 10° as a local baseline, then fit the patch to a Gaussian profile with 7° full width at half maximum. Since the small-scale structure of the maps is dominated by uncorrelated instrument noise, we use a simple Gaussian approximation to the DMR beam instead of the Legendre polynomial expansion discussed in Wright et al. (1994). The pixels are searched in order of decreasing absolute temperature; we subtract each fitted Gaussian from the map before proceeding to the next pixel. To prevent contamination from Galactic plane sources, we only process pixels with Galactic latitude $|b| > 30°$.

The fitted amplitudes $T$ and uncertainties $\delta T$ estimated from the known noise properties of the maps form a data set which we search for evidence of a population of faint unresolved sources. Most of the fitted amplitudes describe noise peaks or the short spatial frequency tail of the scale-invariant CMB temperature distribution. We test this hypothesis and search for any other source population by comparing the fitted amplitudes from the DMR maps to Monte Carlo simulations of instrument noise or a superposition of noise and CMB anisotropy characterized by a Harrison-Zel'dovich spectrum with quadrupole normalization $Q_{rms-PS} = 17$ $\mu$K. We make a cut in signal to noise ratio, $T/\delta T$, and bin the surviving amplitudes to form the integral distribution $N(>T)$ of amplitudes brighter than threshold $T$. Different source populations are expected to have different distributions in temperature. Galactic emission (e.g. from compact HII regions) will appear as pixels of hotter than average temperature, while the Sunyaev-Zel'dovich effect creates a temperature decrement in this wavelength regime. We evaluate separately the integral distributions $N(>T)$ for positive amplitudes and $N(<T)$ for negative amplitudes, as well as the combined distribution $N(>|T|)$ for the absolute value of the fitted amplitude.

Figures 1, 2, and 3 show the integral source distributions $N(>T)$, $N(<T)$, and $N(>|T|)$, respectively, for the DMR 53 GHz data at signal to noise ratio $T/\delta T > 1$, compared to simulations of instrument noise (top) or a superposition of noise and CMB anisotropy (bottom). There are generally more "sources" detected in the (A+B)/2 sum maps than the (A-B)/2 difference maps, consistent with the expected

increase in detections seen in the Monte Carlo simulations when CMB anisotropy is included. We quantify this comparison using the $\chi^2$ statistic

$$\chi^2 = \sum_{i,j} \left(N_i - \langle N_i \rangle\right) \left(\mathbf{M}^{-1}\right)_{ij} \left(N_j - \langle N_j \rangle\right)$$

where $\langle N_i \rangle$ is the mean number of fitted sources in the $i^{\text{th}}$ temperature bin of the simulations and $\mathbf{M}$ is the covariance matrix between bins. We neglect the high-temperature tail of the distribution ($\langle N_i \rangle < 5$) where Poisson statistics dominate, and use an independent test for rare bright sources.

Table 1 shows the $\chi^2$ statistic of the DMR sum and difference maps at signal to noise ratio $T/\delta T > 1$, compared to simulations with noise or a superposition of noise and scale-invariant CMB anisotropy. The probability of obtaining a random realization with $\chi^2$ larger than the DMR value is also shown. We repeat the analysis for several thresholds in signal to noise ratio with generally consistent results. The sum maps have somewhat better $\chi^2$ when compared to simulations which include CMB anisotropy than to a blank sky, but the difference is small, reflecting the low DMR response to multipoles $\ell > 20$. The apparent decrement of negative sources (Figure 2b) is not statistically significant. There is no detection of a population of unresolved sources in the DMR data.

Table 2 compares the total number of positive and negative fitted source amplitudes for the sum and difference maps compared to simulations. There is no evidence for any statistically significant asymmetry between positive and negative amplitudes, in agreement with the lack of statistically significant skewness in the raw DMR pixel temperatures (Smoot et al. 1994). Unresolved Galactic or extragalactic sources, both of which are asymmetric in temperature, contribute negligibly to the anisotropy on 7° scales observed by DMR.

The high-temperature tail of the integral source distribution provides an upper limit to the brightest potential sources in the DMR data. Table 3 shows the brightest fitted source in the sum and difference maps, along with the temperature for which the mean number of sources in the simulated data, plus twice the standard deviation, equals unity. Sources brighter than this limit, should any exist, would be detected. The brightest pixels are in different positions in each map and are consistent with the noise properties of the data. There are no unresolved sources at limit $\Delta T < 179$ $\mu$K (95% CL) antenna temperature in the most sensitive 53 GHz maps.

## 3. Comparison to Known Sources

Bennett et al. (1993) cross-correlate the first-year DMR maps with maps of known non-cosmological sources to place an upper limit on the total contribution of these sources to the anisotropy in the DMR data. We consider here the related question of whether the brightest of these sources are detected in the DMR data, even if the spatial average over all such sources is not.

Inverse Compton scattering of CMB photons from hot intracluster gas causes a decrement in the Rayleigh-Jeans part of the CMB spectrum, $\Delta T = -2y T_0$, where $T_0$ is the unperturbed CMB temperature and the Comptonization parameter $y$ is proportional to the integrated electron pressure along the line of sight (Sunyaev & Zel'dovich 1980). The Sunyaev-Zel'dovich (S-Z) effect has been observed toward a number of X-ray selected clusters (Jones et al. 1993, Grainge et al. 1993, Herbig et al. 1994; see Birkinshaw 1990 for a recent review) with typical amplitude $\Delta T \approx -400~\mu\mathrm{K}$ on arc-minute angular scales. Although beam dilution makes these unlikely to be observed in the DMR data, we have examined the DMR maps for evidence of isolated S-Z sources. Table 4 lists the amplitudes fitted to the DMR sum and difference maps at the positions of firm S-Z detections on arc-minute angular scales (0016+16, A665, A773, and A2218). There is no statistically significant decrement toward these sources.

Nearby clusters are more likely to contribute a detectable signal to the DMR maps given their larger angular extent. We have searched the DMR maps toward nearby clusters at $|b| > 20°$ (Coma, Virgo, Perseus-Pisces, Hercules, and Hydra) for evidence of Sunyaev-Zel'dovich cooling indicative of an extended low surface density cluster atmosphere. To account for diffuse low surface density Comptonization (which may be ill-described by a Gaussian profile), results for these clusters are from a "ring" technique, in which the temperature of all pixels within a disk of radius $\theta$ is compared to the annulus of equal area surronding the disk. There is no detection of the Sunyaev-Zel'dovich cooling as $\theta$ is varied from 5° to 20°. Table 4 lists the results for $\theta = 10°$, placing a limit $\delta y < 7.3 \times 10^{-6}$ (95% CL) for the Comptonization from nearby clusters. Rephaeli (1993) has recently analysed the *HEAO 1*-A2 database to search for evidence of gas associated with superclusters of galaxies. He finds no evidence for such gas, and uses the upper limit on such supercluster gas to infer an upper limit on the S-Z contribution to the DMR maps from such sources of $\delta y < 10^{-7}$. This limit is in agreement with the upper limit $\delta y < 7.3 \times 10^{-6}$ (95% CL) imposed by direct examination of the DMR data toward nearby clusters (Table 4), or the limit $\delta y < 2 \times 10^{-6}$ (95% CL) from the full cross-correlation of the DMR maps with either the rich cluster distribution or the *HEAO 1*-A2 data (Bennett et al. 1993).

Extragalactic radio sources are expected to contribute negligibly to the DMR signal (Franceschini et al. 1989). Table 5 shows the fitted amplitudes towards the positions of the largest apparent diameter objects (the Large Magellanic Cloud and M31) and the brightest sources at 53 GHz from a compilation of radio and millimeter surveys with spectra interpolated to the DMR wavelengths (Witebsky, priv. comm.). The LMC is expected to be the brightest extragalactic source, contributing approximately 50 $\mu$K in a single pixel at 53 GHz (Bennett et al. 1993). The 2$\sigma$ upper limit of 87 $\mu$K from Table 5 is consistent with this level of emission. As expected, fitting individual pixels shows no detection of unresolved extragalactic radio sources in the DMR data.

## 4. Discussion

There is no evidence in the DMR data for unresolved sources beyond the contribution expected from the high spatial frequency tail of a scale-invariant spectrum of anisotropy. The distribution of amplitudes fitted to a Gaussian profile at each pixel is symmetric about zero, implying that unresolved Galactic sources contribute negligibly compared to the CMB and noise in the two-year maps. The high-temperature tail of the fitted amplitude distribution for $|b| > 30°$ is consistent with the noise properties of the data; there are no sources brighter than 179 $\mu$K antenna temperature at 53 GHz (95% CL), corresponding to a flux density 322 Jy.

A search of pixels where extragalactic radio sources are expected to be brightest shows no evidence of emission, nor was any expected. A search toward bright nearby clusters shows no evidence for Sunyaev-Zel'dovich cooling and places a limit on the Comptonization parameter $\delta y < 7.3 \times 10^{-6}$ (95% CL) for these clusters. Since the Sunyaev-Zel'dovich effect is independent of the cluster redshift (except for the angular size), clusters undetectable in the optical or X-ray at high redshifts could in principle affect the DMR maps. The absence of any significant S-Z effect in the DMR maps (e.g. the lack of significant temperature asymmetry) is evidence against significant contamination of the DMR maps from high-redshift clusters.

The recent report of strong unresolved sources at 0.5° angular scale has kindled interest in non-Gaussian models of CMB anisotropy. The MSAM experiment has detected two unresolved sources (width in right ascension $< 15'$) consistent with CMB features of amplitude 4 Jy at 1.8 mm wavelength, or thermodynamic temperature $\Delta T \approx 250~\mu$K (Cheng et al. 1994). These features, if real, may imply a non-Gaussian component to the CMB anisotropy on 0.5° angular scale. Because of beam dilution, the lack of unresolved Planckian sources in the DMR data can not provide meaningful limits to measurements on smaller angular scales. The MSAM sources would contribute less than 2 $\mu$K to the DMR maps, well below the 90 $\mu$K mean noise level per pixel of the DMR 53 GHz sum maps. Conversely, an unresolved source at the DMR upper limit of 322 Jy would dominate a survey at 0.5° angular resolution and so must be rare to avoid detection in surveys of limited sky coverage. If a population of Planck spectrum sources with angular size less than 0.5° were to exist, with integral number density per unit area brighter than flux density $S$ assumed to vary as a power law $n(>S) \sim S^p$, we may normalize the assumed distribution to the MSAM source density (2 sources in 0.002 sr at 4 Jy) and derive an upper limit for larger $S$ based on the DMR results (less than one source in $2\pi$ steradian at 322 Jy) with the result $n(>S) < 2 \times 10^4~(S/1~\mathrm{Jy})^{-2}~\mathrm{sr}^{-1}$. Interpretation of the MSAM measurement or other recent degree-scale anisotropy results (ACME-HEMT, Schuster et al. 1993; MAX, Gundersen et al. 1993; Big Plate, Wollack et al. 1993) in terms of a population of unresolved sources is complicated by the small sampling of the sky (typically 0.1%) and the sparse spatial sampling of the instrumental scan patterns. A CMB survey with 0.5° resolution or better covering a significant fraction of the sky would be immensely valuable regarding the question of whether such sources exist.

We gratefully acknowledge the efforts of those contributing to the *COBE* DMR. Chris Witebsky kindly provided us with compilations of radio sources interpolated to the DMR frequencies and angular resolution. *COBE* is supported by the Office of Space Sciences of NASA Headquarters.

Table 1: Comparison of Fitted Amplitudes to Simulations[a]

| Data Set | Model=Noise + CMB | | Model=Noise Only | |
|---|---|---|---|---|
| | (A+B)/2 | (A-B)/2 | (A+B)/2 | (A-B)/2 |
| 31 GHz $N(>T)$ | 14.7 (18%) | 8.0 (73%) | 15.7 (14%) | 8.5 (66%) |
| 31 GHz $N(<T)$ | 10.3 (50%) | 18.7 ( 7%) | 10.5 (48%) | 20.4 ( 4%) |
| 31 GHz $N(>|T|)$ | 14.2 (29%) | 21.0 ( 5%) | 15.7 (21%) | 22.8 ( 3%) |
| | | | | |
| 53 GHz $N(>T)$ | 7.6 (83%) | 23.7 ( 2%) | 11.5 (38%) | 22.0 ( 2%) |
| 53 GHz $N(<T)$ | 14.0 (29%) | 7.8 (79%) | 11.7 (37%) | 4.3 (96%) |
| 53 GHz $N(>|T|)$ | 11.3 (51%) | 15.4 (20%) | 12.1 (43%) | 9.0 (70%) |
| | | | | |
| 90 GHz $N(>T)$ | 17.0 ( 8%) | 17.6 ( 7%) | 24.9 ( 1%) | 18.7 ( 5%) |
| 90 GHz $N(<T)$ | 3.1 (98%) | 13.6 (19%) | 2.4 (99%) | 15.9 (10%) |
| 90 GHz $N(>|T|)$ | 5.3 (91%) | 14.5 (19%) | 9.2 (61%) | 19.7 ( 4%) |

[a] $\chi^2$ statistic and probability for a random realization to have larger $\chi^2$ than the DMR value. Fitted amplitudes from the DMR maps are compared to Monte Carlo realizations for $|b| > 30°$ and signal to noise ratio $T/\delta T > 1$.

Table 2: Positive and Negative Source Counts in DMR Maps

| Data Set | $N+$[a] | $N-$[a] | $\langle T+ \rangle$[b] ($\mu$K) | $\langle T- \rangle$[b] ($\mu$K) |
|---|---|---|---|---|
| 31 GHz (A+B)/2 | 98 | 109 | 202 | -197 |
| 31 GHz (A-B)/2 | 90 | 107 | 199 | -194 |
| Model (Noise+CMB) | $93 \pm 8$ | $94 \pm 8$ | $201 \pm 7$ | $-202 \pm 7$ |
| Model (Noise Only) | $90 \pm 8$ | $94 \pm 8$ | $202 \pm 7$ | $-202 \pm 7$ |
| | | | | |
| 53 GHz (A+B)/2 | 100 | 99 | 73 | -65 |
| 53 GHz (A-B)/2 | 102 | 92 | 62 | -65 |
| Model (Noise+CMB) | $101 \pm 8$ | $103 \pm 9$ | $70 \pm 2$ | $-70 \pm 2$ |
| Model (Noise Only) | $91 \pm 8$ | $93 \pm 8$ | $69 \pm 2$ | $-68 \pm 2$ |
| | | | | |
| 90 GHz (A+B)/2 | 114 | 86 | 101 | -95 |
| 90 GHz (A-B)/2 | 86 | 102 | 100 | -94 |
| Model (Noise+CMB) | $96 \pm 8$ | $95 \pm 8$ | $97 \pm 3$ | $-96 \pm 3$ |
| Model (Noise Only) | $93 \pm 8$ | $90 \pm 8$ | $97 \pm 3$ | $-95 \pm 3$ |

[a] Total number of sources at $|b| > 30°$ and signal to noise ratio $T/\delta T > 1$.
[b] Mean antenna temperature of fitted source amplitudes.

Table 3: Brightest Fitted Sources in DMR Maps[a]

| Frequency (GHz) | Sum Map ($\mu$K) | Difference Map ($\mu$K) | CMB + Noise Simulation ($\mu$K) | Noise Simulation ($\mu$K) |
|---|---|---|---|---|
| 31 | 531 | 502 | 527 | 542 |
| 53 | 171 | -138 | 179 | 171 |
| 90 | 206 | -266 | 241 | 236 |

[a] Fitted amplitudes at $|b| > 30°$ and signal to noise ratio $T/\delta T > 1$. Amplitudes are antenna temperature.

Table 4: Fitted Amplitudes Toward Sunyaev-Zel'dovich Candidates ($\mu$K)

| Source | $T_{31}$ | $T_{53}$ | $T_{90}$ | Weighted Mean |
|---|---|---|---|---|
| 0016+16[a] | -40 ± 182 | -7 ± 60 | -61 ± 98 | -23 ± 49 |
| | -46 ± 182 | -12 ± 60 | -47 ± 98 | -23 ± 49 |
| A665[a] | -29 ± 120 | 12 ± 48 | -5 ± 78 | 3 ± 38 |
| | 68 ± 120 | 34 ± 48 | -8 ± 78 | 27 ± 38 |
| A773[a] | 31 ± 138 | -13 ± 52 | -58 ± 85 | -20 ± 42 |
| | -13 ± 138 | 24 ± 52 | 0 ± 85 | 15 ± 42 |
| A2218[a] | -179 ± 115 | 12 ± 45 | 28 ± 74 | -3 ± 36 |
| | -38 ± 115 | -12 ± 45 | -10 ± 74 | -14 ± 36 |
| Coma[b] | 36 ± 67 | 10 ± 23 | -19 ± 37 | 5 ± 19 |
| | -102 ± 67 | -14 ± 23 | 16 ± 37 | -13 ± 19 |
| Virgo[b] | 26 ± 76 | -48 ± 24 | 26 ± 39 | -24 ± 20 |
| | 60 ± 76 | 13 ± 24 | -44 ± 39 | 2 ± 20 |
| Perseus-Pisces[b] | 112 ± 87 | 29 ± 34 | 27 ± 54 | 37 ± 27 |
| | -2 ± 87 | 3 ± 34 | -15 ± 54 | -2 ± 27 |
| Hercules[b] | 39 ± 59 | -15 ± 25 | -11 ± 40 | -8 ± 20 |
| | -117 ± 59 | 11 ± 25 | 59 ± 40 | 8 ± 20 |
| Hydra[b] | -137 ± 83 | -21 ± 29 | -19 ± 47 | -30 ± 24 |
| | 12 ± 83 | -5 ± 29 | -16 ± 47 | -6 ± 24 |

[a] First line for each source refers to the fitted amplitudes in the (A+B)/2 sum map, while the second line refers to the (A-B)/2 difference map. All amplitudes are thermodynamic temperature.

[b] Thermodynamic temperature differences from "ring" technique.

Table 5: Fitted Amplitudes Toward Radio Sources ($\mu$K)

| Source | $T_{31}$ | $T_{53}$ | $T_{90}$ |
|---|---|---|---|
| LMC[a] | 38 ± 106 | 11 ± 38 | 0 ± 55 |
| | 125 ± 106 | 3 ± 38 | 31 ± 55 |
| M31 | 92 ± 162 | 61 ± 57 | -7 ± 82 |
| | 63 ± 162 | 2 ± 57 | -80 ± 82 |
| 3C273.0 | 0 ± 189 | 11 ± 53 | -8 ± 77 |
| | -30 ± 189 | 21 ± 53 | -19 ± 77 |
| 3C345 | -42 ± 100 | 33 ± 37 | -6 ± 54 |
| | -19 ± 100 | -27 ± 37 | 8 ± 54 |
| 0537-441 | 12 ± 100 | 0 ± 33 | -1 ± 47 |
| | 71 ± 100 | 6 ± 33 | 39 ± 47 |

[a] First line for each source refers to the fitted amplitudes in the (A+B)/2 sum map, while the second line refers to the (A-B)/2 difference map. All amplitudes are antenna temperature.

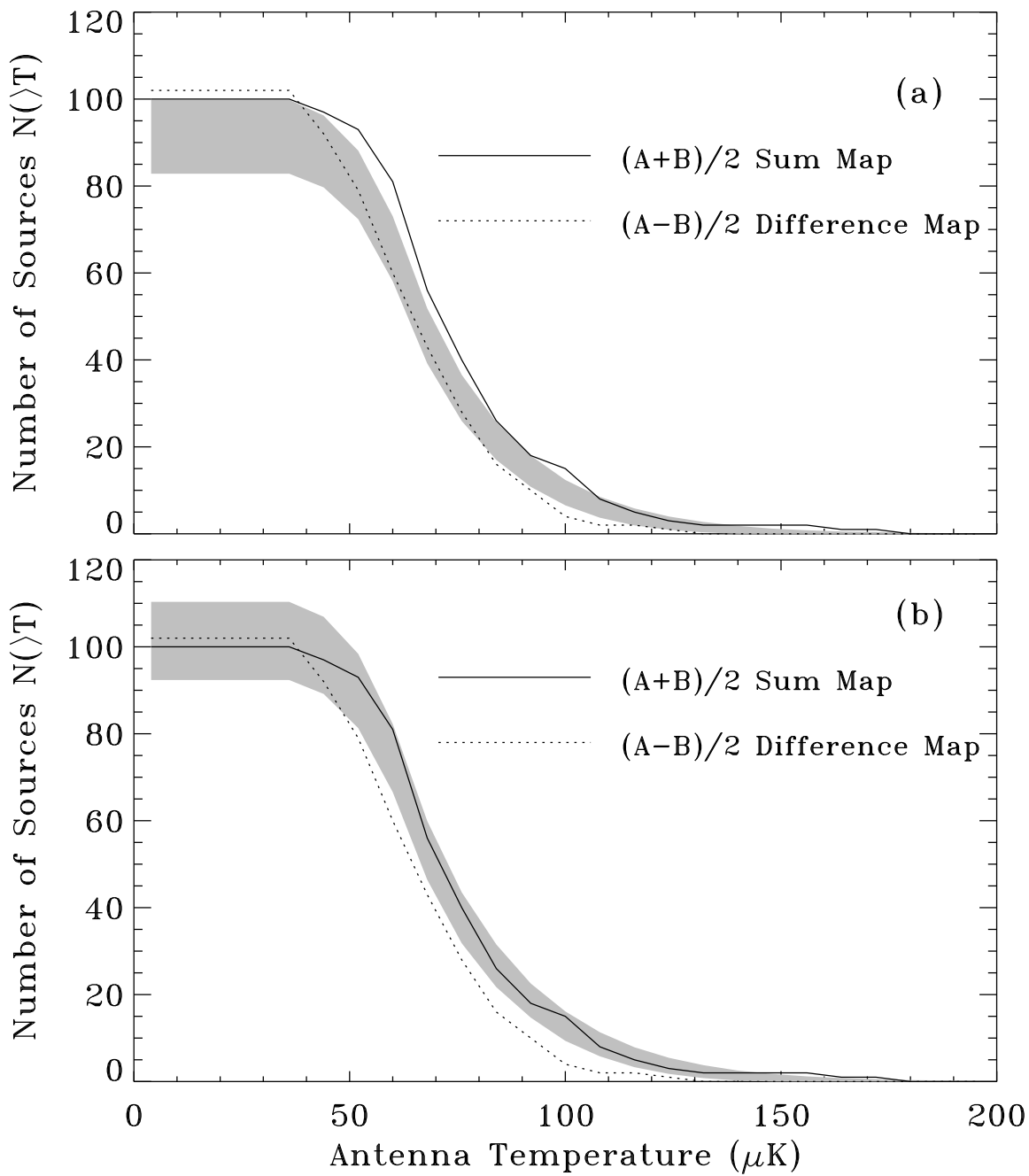

Fig. 1.— Integral source counts $N(>T)$ for the DMR 53 GHz (A+B)/2 sum maps and (A-B)/2 difference maps at $|b| > 30°$ and signal to noise ratio $T/\delta T > 1$. The cut in signal to noise accounts for the flattening of the curves near zero amplitude. (a) Gray band is 68% confidence range of simulations of instrument noise. (b) Gray band is 68% confidence range of simulations of a superposition of instrument noise and scale-invariant CMB anisotropy normalized to $Q_{rms-PS} = 17 \mu K$.

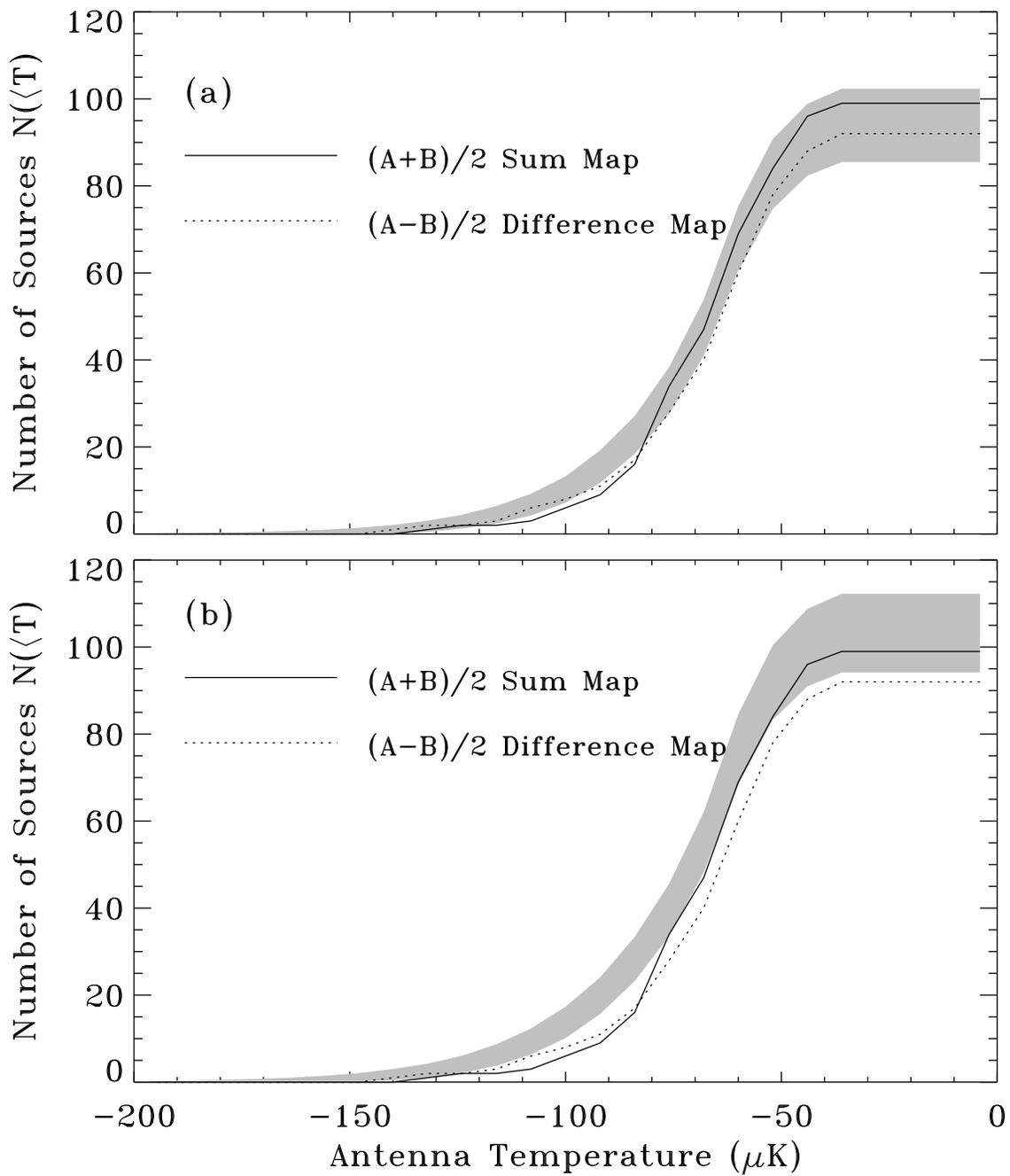

Fig. 2.— Integral source counts $N(<T)$ at 53 GHz. Same parameters as Figure 1.

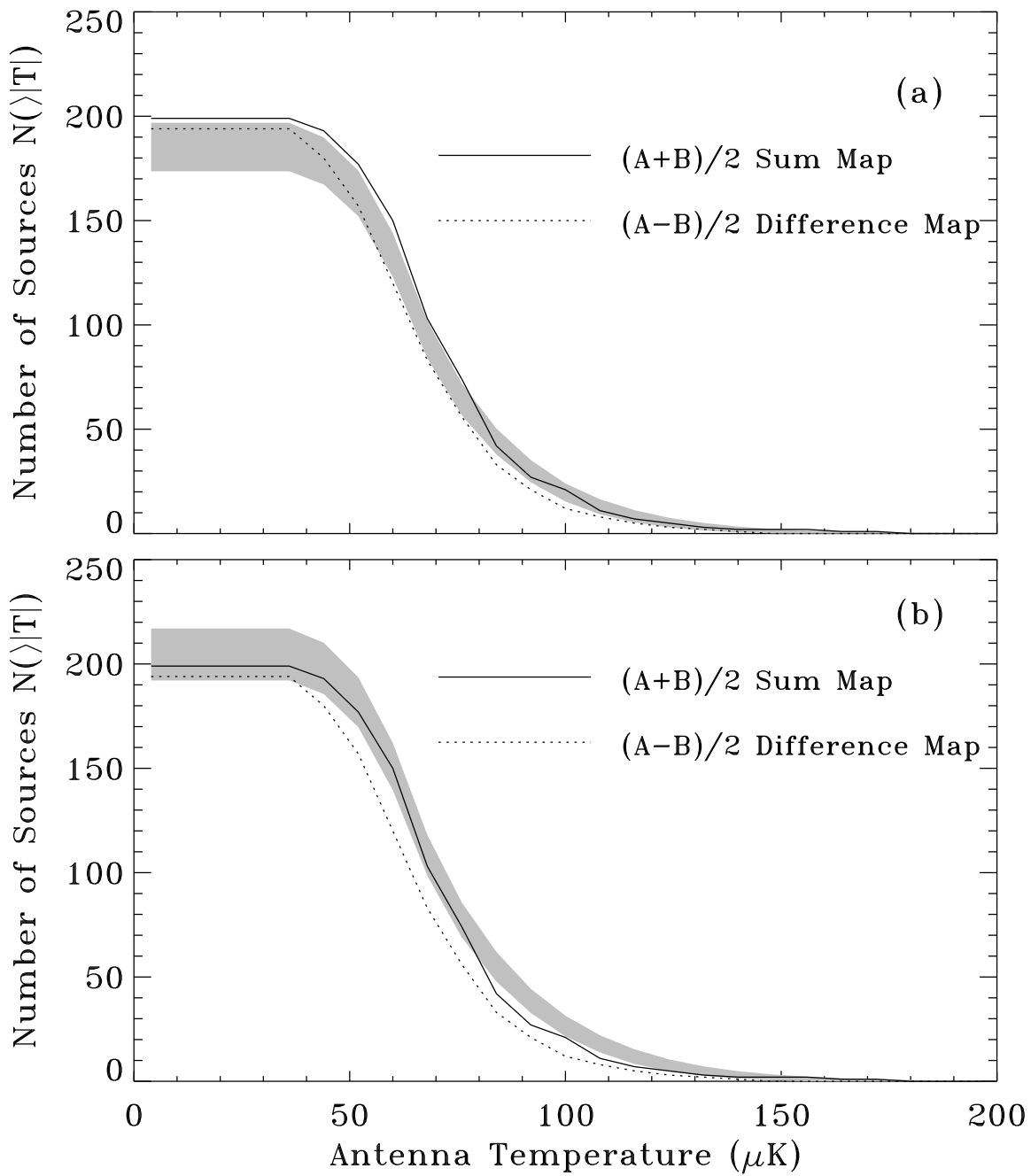

Fig. 3.— Integral source counts $N(>|T|)$ at 53 GHz. Same parameters as Figure 1.